\documentclass[useAMS,usenatbib]{mn2e}

\usepackage{graphicx}
\usepackage{color}
\usepackage{amssymb}
\usepackage{arydshln}

\title[Degeneracy in dwarf galaxy simulations]
  {The degeneracy between star-formation parameters in dwarf galaxy
   simulations and the $M_{\rm star}-M_{\rm halo}$ relation}
\author[A. Cloet-Osselaer et al.]
  {A.~Cloet-Osselaer$^1$\thanks{ACO and SDR thank the Ghent University
  Special Research Fund for financial support, E-mail:
  Annelies.Cloet-Osselaer@UGent.be.}, S.~De Rijcke$^1$,
  J.~Schroyen$^1$\thanks{thanks the Fund for Scientific Research -
  Flanders, Belgium (FWO), E-mail: Joeri.Schroyen@UGent.be.},  V.~Dury$^1$\thanks{thanks the Fund for Scientific Research -
  Flanders, Belgium (FWO)}
  \\ $^1$Sterrenkundig Observatorium,
  Ghent University, Krijgslaan 281, S9, 9000 Gent, Belgium }
\date{Accepted , Received ; in original form}

\pagerange{\pageref{firstpage}--\pageref{lastpage}} \pubyear{2011}

\def\LaTeX{L\kern-.36em\raise.3ex\hbox{a}\kern-.15em
    T\kern-.1667em\lower.7ex\hbox{E}\kern-.125emX}

\begin{document}

\label{firstpage}

\maketitle

\begin{abstract}
We present results based on a set of N-Body/SPH simulations of isolated dwarf galaxies. The simulations take into account star formation, stellar feedback, radiative cooling and metal enrichment. 
The dark matter halo initially has a cusped profile, but, at least in these simulations, starting from idealised, spherically symmetric initial conditions, a natural conversion to a core is observed due to gas dynamics and stellar feedback.

A degeneracy between the efficiency with which the interstellar medium absorbs energy feedback from supernovae and stellar winds on the one hand, and the density threshold for star formation on the other, is found. We performed a parameter survey to determine, with the aid of the observed kinematic and photometric scaling relations, which  combinations of these two parameters produce simulated galaxies that are in agreement with the observations. 

With the implemented physics we are unable to reproduce the relation between the stellar mass and the halo mass as determined by \cite{guo10}, however we do reproduce the slope of this relation.
\end{abstract}


\section{Introduction}

Dwarf galaxies are the most common type of galaxy in the local
universe but also the faintest and least easy to observe.  In the
$\Lambda$CDM cosmology, our universe consists of matter, both
luminous and dark, and dark energy, which is responsible for the
accelerating expansion of the universe. Galaxies form when gas
collapses in dark matter halos. Baryons, be it in the form of gas, dust
or stars, are the most accessible form of matter, emitting radiation
over the whole electromagnetic spectrum. Dark matter, on the other
hand, as it only interacts gravitationally, is much more difficult to
``observe''.

There have been many attempts to estimate dark halo masses and
mass-to-light ratios for galaxies and clusters of galaxies from direct
observations. These include methods that make use of gravitational
lensing \citep{mandelbaum06, liesenborgs09},
dynamical modeling of the observed properties of a kinematical tracer
such as stars or planetary nebulae \citep{kronawitter00, derijcke06, napolitano11, barnabe09}.
One thing virtually all these works have in
common is the relatively limited size of the data set they are based
on. \cite{guo10} determined the halo mass as a function of stellar
mass for a large sample of galaxies using a statistical analysis of
the Sloan Digital Sky Survey, which yields the stellar masses, and the
Millennium Simulations, which yield the dark-matter masses. In the
range of the most massive halos and bright galaxies, the derived
$M_{\rm star}$-$M_{\rm halo}$ relation, which is of the form $M_{\rm
  star} \propto M_{\rm halo}^{0.36}$, is found to be in good
agreement with gravitational lensing data \citep{mandelbaum06}. Below
a halo mass of $M_{\rm halo} \sim 10^{11.4}$~$M_\odot$, this relation
becomes much steeper:~$M_{\rm star} \propto M_{\rm
  halo}^{3.26}$. \cite{guo10} extrapolate the latter relation into the
dwarf regime, where $M_{\rm halo} \lesssim 10^{10}$~$M_\odot$. This
leads then to the prediction that faint dwarf galaxies with stellar
masses of the order of $M_{\rm star} \sim 10^6$~$M_\odot$ should live
in comparatively  massive $M_{\rm halo} \sim 10^{10}$~$M_\odot$
dark-matter halos.

The \cite{guo10} $M_{\rm star}$-$M_{\rm halo}$ relation was compared
with that found in simulations of dwarf galaxies 
\citep{valcke08,stinson07,stinson09,governato10,pelupessy04,mashchenko08}
by \cite{sawala11a} and \cite{sawala11b}. They found that simulated
dwarf galaxies had stellar masses that were at least an order of
magnitude higher at a given halo mass than predicted by
\cite{guo10}. There could be several causes for numerical dwarf
galaxies to be overly prolific star formers:
\begin{itemize}
\item The star formation efficiency could be too high because of an
  underestimation of the feedback efficiency. \cite{stinson06}
  investigated the influence of the feedback efficiency on the mean
  star formation rate (SFR). The general trend they have observed was a
  decrease of the mean SFR when increasing the feedback efficiency.
\item \cite{stinson06} also reported finding a decreasing mean SFR with
  increasing density threshold for star formation. Recently, high
  density thresholds for star formation have come in vogue, see
  e.g. \cite{governato10}.
\item Dwarf galaxies, due to their low masses, are expected to be
  particularly sensitive to reionisation. Not properly taking into
  account the effects of reionisation may lead to an overestimation of
  the gas content of dwarfs and an underestimation of the gas cooling
  time.
\item Dwarf galaxies are metal poor and hence also dust poor. This
  lowers the production of H$_2$ molecules and causes poor
  self-shielding of molecular clouds \citep{buyle06} which
  could be expected to inhibit star formation. Not taking these
  effects into account will lead to an overestimation of the
  SFR \citep{gnedin09}.
\end{itemize}

Using the high values for the density threshold above which gas
particles become eligible for star formation, denoted by $\rho_{\rm SF}$, as promoted by \cite{governato10}, in combination with
radiative cooling curves that allow the gas to cool below $10^4$~K
\citep{maio07}, makes the gas collapse into small, very dense and cool clouds
before star formation ignites. If the supernova feedback
$\epsilon_{\rm FB}$, defined as the fraction of the average energy
output of a supernova that is actually absorbed by the interstellar
medium (ISM), is too weak to sufficiently heat and/or disrupt such a
star-forming cloud, one can consequently expect the mean SFR to be
very high, leading to overly massive (in terms of $M_{\rm star}$)
dwarfs. Therefore, one could hope to remedy this situation by
increasing $\epsilon_{\rm FB}$ accordingly. In that case, a
correlation between $\epsilon_{\rm FB}$ and $\rho_{\rm SF}$ would be
expected to exist.

In the present paper, we analyze a large suite of numerical
simulations of isolated, spherically symmetric dwarf galaxies in which
we varied both the feedback efficiency $\epsilon_{\rm FB}$ and the
density threshold $\rho_{\rm SF}$. Our goal is to investigate {\em (i)}
if such a correlation between $\epsilon_{\rm FB}$ and $\rho_{\rm SF}$
exists and, if it exists, how to break it, {\em (ii)} which
$\epsilon_{\rm FB}$/$\rho_{\rm SF}$-combinations lead to viable dwarf
galaxy models in terms of the observed photometric and kinematic
scaling relations, and {\em (iii)} how well these models approximate
the aforementioned $M_{\rm star}$ -$M_{\rm halo}$ relation.

In section \ref{section:numerical_details}, we give more details about the numerical methods that
are used in our code. An analysis of the simulations is given in
section \ref{section:analysis}, where some details are given of the NFW halo that is used
for the simulations and a large set of scaling relations are
plotted comparing our models to observations. In section \ref{section:results} we discuss the obtained results and
 conclude.


\section{Numerical details}
\label{section:numerical_details}

We use a modified version of the Nbody-SPH code {\sc Gadget-2}
\citep{springel05}. The original {\sc Gadget-2} code was extended with
star formation, feedback and radiative cooling by \cite{valcke08}.
While the initial conditions of the simulations are cosmologically
motivated (see below), we do not perform full cosmological
simulations. Our approach yields a high mass resolution at
comparatively low computational cost. Still, previous work by
\cite{valcke08}, \cite{valcke10} and \cite{schroyen11} has shown that
with this code realistic dwarf galaxies, following the known
photometric and kinematic scaling relations, can be produced.
We set up the simulations using 200,000 gas particles and 200,000 DM particles.
Depending on the model's total mass, this results in gas particle masses in the range of 
$350-2,620 M_{\odot}$ and DM particle masses in the range of $1,650-12,380 M_{\odot}$. 
We use a gravitational softening length of 0.03~kpc.

Our results are visualized with our own software package HYPLOT. This
is freely available from SourceForge\footnote{http://sourceforge.net/projects/hyplot/}
 and is used for all the figures
in this paper.

\subsection{Initial conditions}

Our models are set up, as in \cite{valcke08,valcke10,schroyen11}, with a spherically symmetric dark matter halo
and a homogeneous gas cloud. This gas cloud has a density of $5.55~\rho_{\rm crit}$, with $\rho_{\rm crit}$ the critical density of the universe at the halo's formation redshift, here taken to be $z_{\rm c}=4.3$. This is equivalent with a number density for the gas of 0.0011 hydrogen atoms per cubic centimeter.
We use a flat $\Lambda$-dominated cold dark matter cosmology with the following cosmological parameters: 
$h = 0.71, \Omega_{\rm tot} = 1, \Omega_{\rm m} = 0.2383, \Omega_{\rm DM} = 0.1967$. The baryonic mass fraction will be the difference between $\Omega_{\rm m}$ and $\Omega_{\rm DM}$, in practice it will have a value that is 0.2115 times that of the dark-matter.
At the start of the simulations the gas particles are initially at rest, their initial metallicities are set to $10^{-4}\ Z_{\odot}$ and their initial temperature is $10^4$ K.
The dark matter halo has a NFW density profile \citep{NFW}:
\begin{equation}
 \rho_{\rm NFW}(r) = \frac{\rho_{\rm s}}{(r/r_{\rm s})(1+r/r_{\rm s})^{2}}
\end{equation}
where $\rho_{\rm s}$ and $r_{\rm s}$ are, respectively, the
characteristic density and the scale radius. In order to fix the values
of these parameters, we use the correlation between them found by
\cite{wechsler02} and \cite{gentile04}, which makes the NFW density
distribution essentially a one-parameter family of the dark matter
virial mass, $M_{\rm DM}$. The relations we use for $\rho_{\rm s}$,
$r_{\rm s}$ and the concentration parameter $c$ (=$r_{\rm max}/r_{\rm
  s}$) are~:
\begin{eqnarray}
 c &\simeq& 20 \left(\frac{M_{\rm DM}}{10^{11}M_{\sun}}\right)^{-0.13}
 \\ r_{\rm s} &\simeq& 5.7 \left(\frac{M_{\rm
     DM}}{10^{11}M_{\sun}}\right)^{0.46}~{\rm kpc} \\ \rho_{\rm s}
 &\simeq& \frac{101}{3}\frac{c^{3}}{\ln(1+c)-c/(1+c)}\rho_{\rm crit}.
\end{eqnarray}
Here, $r_{\rm max}$ is the halo's
virial radius. At $r_{\rm max}$, the DM halo is truncated and the
density drops to zero, so the entire mass $M_{\rm DM}$ is situated
inside the radius $r_{\rm max}$.

\subsection{Criteria for Star formation}
\label{subsection:SFC}
Star formation is assumed to take place in cold, dense, converging and
gravitationally unstable molecular clouds \citep{katz96}. Gas
particles that fulfill the star formation criteria (SFC) are eligible
to be turned into stars. These SFC are:
\begin{eqnarray}
 \rho_{\rm g} & \geq & \rho_{\rm SF} \\ T & \leq & T_{\rm c} = 15000 \mathrm{K}
 \\ \vec{\nabla}.\vec{v} & \leq & 0,
\end{eqnarray}
with $\rho_{\rm g}$ the gas density, $T$ its temperature and $\vec{v}$
its velocity field. $\rho_{\rm SF}$ is the density threshold for star
formation. We employ a Schmidt law \citep{schmidt59} to convert gas
particles that fulfill the SFC into stars:
\begin{equation}
\frac{\mathrm{d}\rho_{\rm s}}{\mathrm{d}t} =
-\frac{\mathrm{d}\rho_{\rm g}}{\mathrm{d}t} =
c_{\star}\frac{\rho_{\rm g}}{t_{\rm g}}, \label{cstar}
\end{equation}
with $\rho_{\rm s}$ the stellar density and $c_\star$ the
dimensionless star formation efficiency. The timescale $t_{\rm g}$ is
taken to be the dynamical time for the gas $1/\sqrt{4 \pi G \rho_{\rm
    g}}$. Here, we choose $c_{\star}=0.25$. \cite{stinson06} showed that
the influence on the mean SFR of the value of $c_{\star}$ with values in
the range of 0.05 to 1 is negligible. Lowering $c_{\star}$ reduces the
star formation efficiency as well as the amount of supernova feedback, causing
more particles to fulfill the density and temperature criteria. This
compensates for the lower value of $c_{\star}$, producing a SFR which is
roughly independent of $c_\star$. 

\cite{revaz09} also investigated the influence of $c_\star$ by varying it between the values of 0.01 and 0.3. They concluded that the star formation history is mainly determined by the initial total mass with a minor influence of $c_\star$. Self-regulating models, in which star formation occurs in recurrent bursts due to the interplay between cooling and supernova feedback, were achieved for $c_{\star}\sim0.2$. 
            Such models best resemble real dwarf galaxies.

\subsection{Feedback}
We consider feedback from star particles by supernova Ia (SNIa),
supernova II (SNII) and stellar winds (SW). They deliver energy and
mass to the ISM and enrich the gas. Feedback is
distributed over the gas particles in the
neighborhood of the star particle according to the SPH smoothing kernel. Each star particle represents a
single-age, single-metallicity stellar population (SSP). The stars within each SSP
are distributed according to a Salpeter initial mass function~:
 \begin{equation}
\Phi(m)\mathrm{d}m = Am^{-(1+x)}\mathrm{d}m,
\end{equation}
with $x=1.35$ and $A=0.06$. The limits for the stellar masses are
$m_{\mathrm{l}} = 0.01~\mathrm{M_{\odot}}$ and $ m_{\mathrm{u}} =
60~\mathrm{M_{\odot}}$. The energy release of a SN is set to $E_{\rm
  tot} = 10^{51}$~erg and that by a SW to $E_{\rm tot} = 10^{50}$~erg
\citep{thornton98}. The actual energy injected into the ISM is
implemented as $\epsilon_{\rm FB} \times E_{\rm tot}$, where
$\epsilon_{\rm FB}$ is a free parameter.


\subsection{Cooling}
Metallicity-dependent radiative cooling is implemented using the cooling curves 
from \cite{sutherland93}. With this recipe it is possible
to cool gas to a minimum temperature of $10^{4}$~K. 
We also implemented the \cite{maio07} cooling curves, making it possible for particles to cool below $10^{4}$~K.

\subsection{Production runs}

In Table \ref{Models_table}, we give an overview of the parameters
that were used to set up the models. A benefit of our code is that we can retain the same initial conditions
and easily adapt our parameters to perform a detailed parameter
survey. In the remainder, we will quantify the density threshold by $n_{\rm SF}$ expressed in hydrogen ions 
per cubic centimeter (so $\rho_{\rm SF} = 1~{\rm amu} \times n_{\rm SF}$). At the start of the simulations, 
the models only contain
gas and dark matter. During the first few $10^8$~years, the gas collapses
in the gravitational potential well of the DM. The simulations run for
12.22 Gyr, till $z=0$. 

\begin{table}
 \caption{Details of the basic spherical dwarf galaxy models that were
   used in the simulations. Initial masses for the DM halo and gas
   are given in units of $10^{6} \mathrm{M_{\odot}}$, radii in
   kpc.\label{Models_table}} 
\begin{tabular}{lcccc}
\hline                                
$\mathrm{model}$ & $M_{\rm DM,i}$ & $M_{\rm g,i}$ & $r_{\rm s}$ & $r_{\rm max}$\\ 
\hline
N03 & 330 & 70 & 0.412 & 17.319\\ 
N05 & 660 & 140 & 0.566 & 21.742\\
N06 & 825 & 175 & 0.627 & 23.393\\ 
N07 & 1238 & 262 & 0.756 & 26.755\\
N08 & 1654 & 349 & 0.863 & 29.428\\ 
N09 & 2476 & 524 & 1.040 & 33.634\\ 
 
\hline 
\end{tabular}

\end{table}

In the literature, a large variety of values for the density threshold can be found. \cite{stinson06} use a low density 
threshold of $0.1$~cm$^{-3}$ while \cite{governato10} use a high density threshold of $100\ \rm cm^{-3}$ which, these authors
argue, is a better representation of the conditions in star-forming regions in real galaxies. The simulations of \cite{sawala11a} 
have been performed with a density threshold of $10\ \rm cm^{-3}$. In this paper, we increase the density threshold from $n_{\rm SF}=0.1$~cm$^{-3}$, over 
$n_{\rm SF}=6$~cm$^{-3}$ to $n_{\rm SF}=50$~cm$^{-3}$. For the fiducial series of low-density threshold simulations, we matched the $n_{\rm SF}=0.1$~cm$^{-3}$ 
with a feedback efficiency of $\epsilon_{\rm FB}=0.1$. For the intermediate-density threshold simulations, with $n_{\rm SF}=6$~cm$^{-3}$, we varied the 
feedback efficiency between $\epsilon_{\rm FB}=0.1$ and $0.9$. Finally, for the high-density threshold simulations, with $n_{\rm SF}=50$~cm$^{-3}$, we varied the 
feedback efficiency between $\epsilon_{\rm FB}=0.3$ and $0.9$.



\section{Analysis}
\label{section:analysis}

\subsection{The NFW halo}
\label{section_nfw_stable}

\begin{figure}
\centering \includegraphics[width=0.48\textwidth]{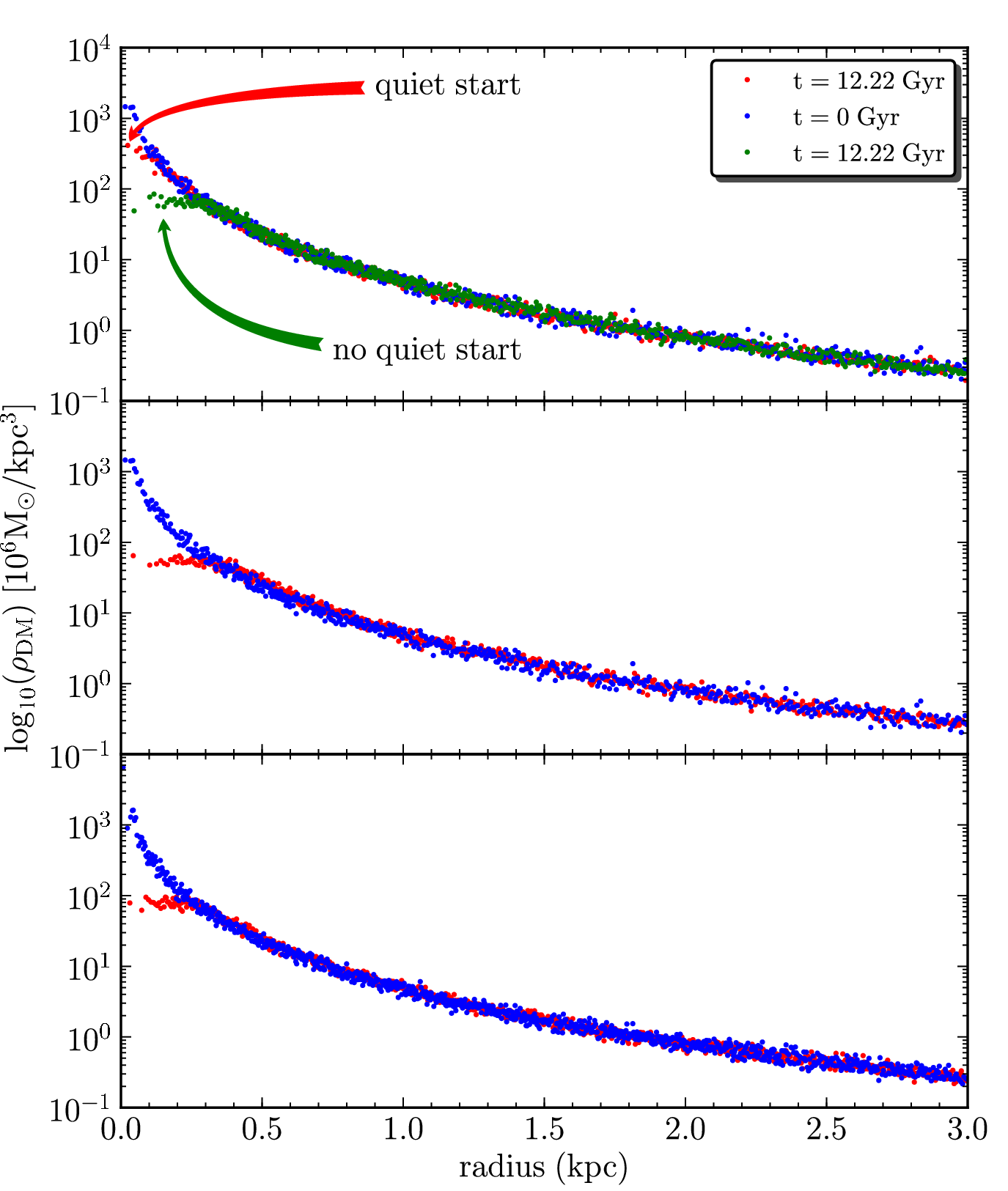}
 \caption{The density profile of the N03 NFW halo for different
   simulations: in the upper panel only DM was included, in the central panel
   DM, gas is included but star formation was turned off. The bottom
   panel shows the results of a simulation with DM, gas and star
   formation. \label{fig:NFWstability}}
\end{figure}

The DM halo is constructed using a Monte Carlo sampling
technique. First, for each particle, the three position coordinates in
spherical coordinates ($r,\ \theta,\ \phi$) are generated.  $r$ is
drawn from the density profile $\rho_{\rm NFW}$ using a standard
acceptance-rejectance technique, $\phi$ and $\cos(\theta)$ are drawn
from uniform distributions over the intervals $[0,2\pi]$ and $[-1,1]$,
respectively. Next, $v_{\rm r}$, $v_{\rm \theta}$ and $v_{\rm \phi}$
are drawn from the isotropic distribution function for the NFW model,
again with an acceptance-rejectance technique. This isotropic
distribution function was constructed from the NFW density profile
using the standard Eddington formula \citep{buyle07}. For each
particle, a symmetric partner was constructed with position
coordinates $(r,\ -\theta,\ -\phi)$ and velocity coordinates $(-v_{\rm
  r}, -v_{\rm \theta}, -v_{\rm \phi})$. This drastically improved the
stability of the central parts of the halos. The very inner part of
the steep cusp of the NFW model is populated by relatively few
particles, destroying its spherical symmetry and introducing
unbalanced angular momenta. This initial deviation leads to the
ejection of particles from the cusp and triggers a more widespread
dynamical response of the DM halo, over time erasing the inner
cusp. Introducing the partner particles, cancelling out the angular
momenta and increasing the symmetry of the particles' spatial
distribution, greatly alleviates these problems. Such techniques for
constructing ``quiet'' initial conditions have been applied before
with great success, see e.g. \citep{selwood86}. The
  improvement of the stability of the DM halo in simulations with a
  ``quiet'' start over simulations without a ``quiet'' start is
  illustrated in the top panel of Fig.  \ref{fig:NFWstability} where
  the density distribution of both haloes at $z=0$ is plotted as red
  and green dots, respectively.

First, to test the stability of the NFW halos, we ran several simulations for the N03 and N05 mass models:
\begin{description}
 \item[\textbf{Run 1:}] only DM
 \item[\textbf{Run 2:}] DM and gas but no star formation
 \item[\textbf{Run 3:}] DM and gas and star formation
 \end{description}
For these test simulations an $n_{\rm SF}$ of $0.1\ \rm cm^{-3}$
\citep{katz96} and $\epsilon_{\rm FB}$ of $0.1$ \citep{thornton98} was
used. 

Fig. \ref{fig:NFWstability} shows the density profile of the test
simulations for the N03 mass model. From the upper panel, it is evident that the DM density of the 
DM-only simulation remains stable and cusped until
the end of the simulation. The simulations presented in the middle and bottom panels, 
show a clear conversion of the cusp into a core over time. Moreover, the width of the core depends on the mass of
the system, with more massive halos having larger cores. 

Our simulations largely confirm the results from \cite{read05}, where a rapid removal of gas results in a conversion from cusp to core as stated first by \cite{navarro96}.
As gas cools and flows into the halo, the center of the dark matter
halo is adiabatically compressed. Without star formation, the central gas pressure builds up,
eventually stops further inflow, and even makes the gas re-expand somewhat. This re-expansion
happens rapidly enough for the DM halo to respond non-adiabatically:~the central DM density 
experiences a net lowering and the cusp is transformed into a core. With star formation turned on,
feedback is responsible for a fast removal of gas from the central parts of the
DM halo, with the same effect:~a conversion from a cusp to a core. 

Unlike us, \cite{governato10} found that the density threshold
  for star formation needed to be high enough for a cusp-to-core
  conversion to occur. Only for $n_{\rm SF} \gtrsim 10$~cm$^{-3}$ does
  supernova feedback lead to sufficient gas motions to flatten the
  cusp in their simulated dwarfs, which are taken from a larger
  cosmological simulation. In contrast, in our more idealized,
  initially spherically symmetric setup, even a low density threshold
  leads to sufficient gas outflow for the cusp to flatten.  

\subsection{Star formation histories}
\label{section_SFH}

\begin{figure}
 \centering \includegraphics[width=0.5\textwidth]{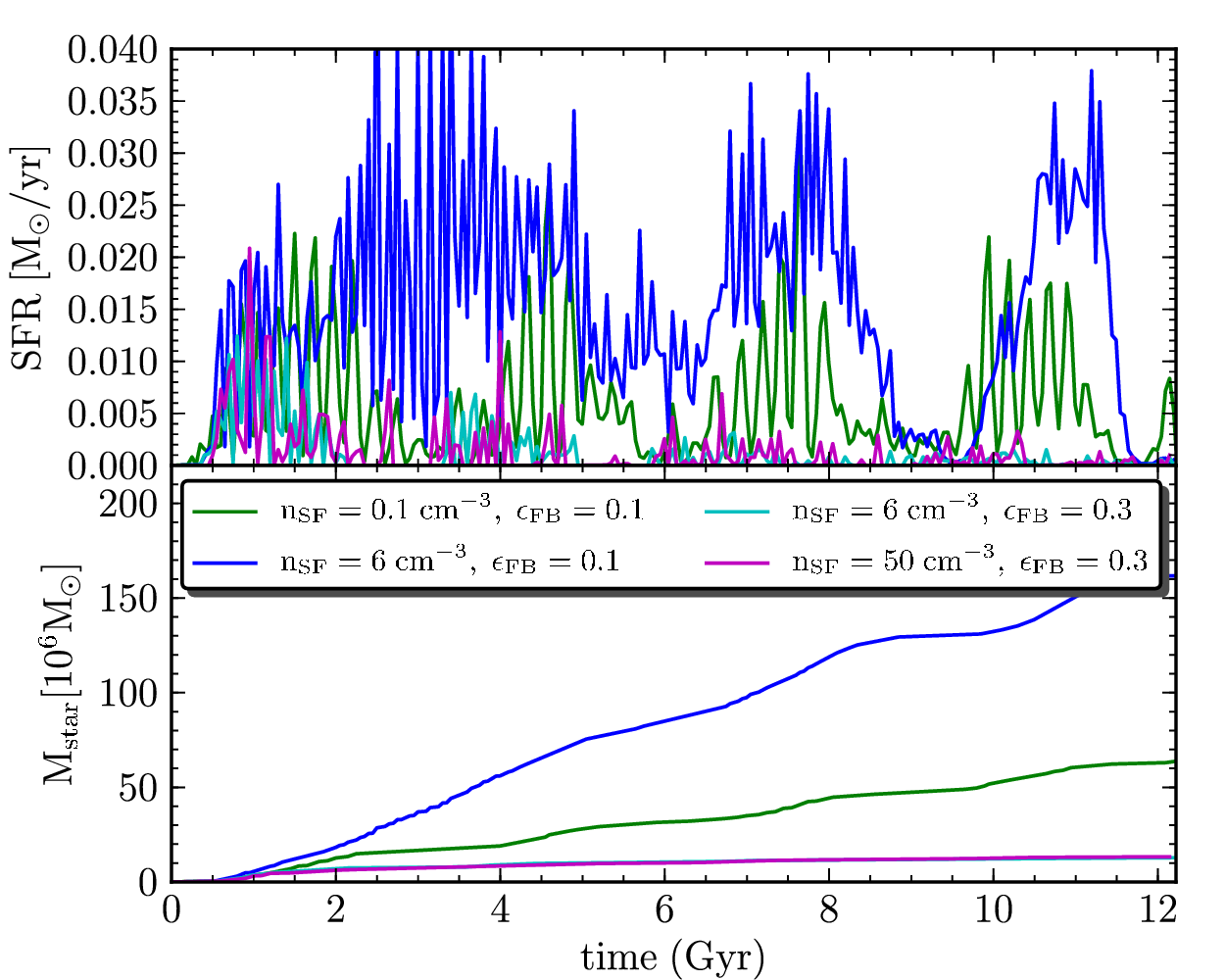}
 \caption{Top panel:~the SFR of several N07 models as a function of time. Bottom panel:~the stellar
   mass as a function of time.\label{fig:SFR}}
\end{figure}

In Fig. \ref{fig:SFR}, we show the star-formation histories (SFHs) of 
different realizations of the N07 mass model. Also, in table \ref{table_finalprop}, the starting time of star formation is tabulated along with the final total stellar mass. Several conclusions can be drawn:
\begin{itemize}
 \item The delay between the start of the simulation and the start of the 
first star-formation event is an increasing function of $n_{\rm SF}$. This appears 
logical:~it takes longer for the gas to collapse to higher densities and ignite star formation.
Comparing different mass models, star formation starts earlier in more massive models for a given 
$n_{\rm SF}$. This is most likely due to the more massive models having steeper gravitational
potential wells, increasing their ability to compress the inflowing gas.
\item If $n_{\rm SF}$ is increased while  $\epsilon_{\rm FB}$ is kept fixed, more stars are formed 
(e.g. going from the green to the blue curve or similarly from the cyan to the magenta curve in Fig.  \ref{fig:SFR}). This is because gas collapses
to higher densities and the feedback is no longer able to sufficiently heat and expel this gas and to interrupt star formation.
\item Related to the previous point, the SFR also becomes more rapidly varying if $n_{\rm SF}$ is increased while 
$\epsilon_{\rm FB}$ is kept fixed. The reason is that in the small high-density star-forming regions, feedback can only 
locally interrupt star formation during short timespans. At lower $n_{\rm SF}$, star formation is more widespread, leading
to more global behavior:~as supernovae go off, star formation can be completely halted.
\item Increasing $\epsilon_{\rm FB}$ while $n_{\rm SF}$  is kept fixed leads to a decrease in star formation (e.g. going from the blue
to the cyan curve in Fig. \ref{fig:SFR}). This is because once feedback is strong enough, it is able to extinguish 
star formation, even at high gas densities.
\item The most low-mass models fail to form stars for high $n_{\rm SF}$ values. E.g. 
no stars form in the N03 models for $n_{\rm SF}>0.1$~cm$^{-3}$. This is due to the masses of these models being too small for gas to collapse to densities where stars can be formed. This point is further elaborated in the next paragraph.
\end{itemize}

\subsection{Density distribution of the ISM}
\label{section_ISM}

\begin{figure}
 \centering \includegraphics[width=0.5\textwidth]{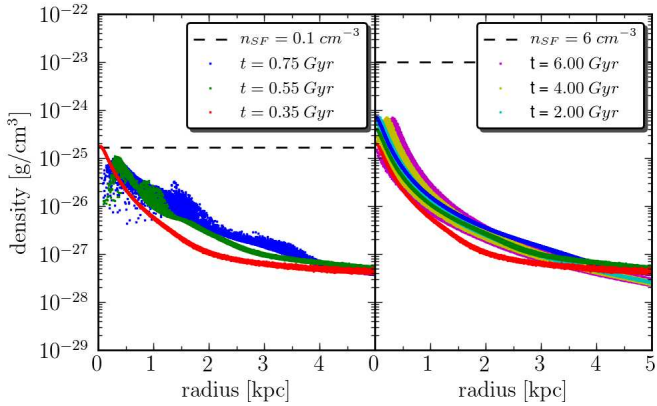}
 \caption{The density distribution of the ISM at different times for the least massive galaxy, N03, with different density threshold and a fixed feedback efficiency of 0.1.\label{fig:N03}}
\end{figure}

In Fig. \ref{fig:N03}, the density of the ISM is plotted as a function of radius. For the N03 model in the left panel a density threshold of $0.1$~cm$^{-3}$ was used while for the model in the right panel, the density threshold was set to a value of $6$~cm$^{-3}$. The red points show the gas distribution at the moment just before the start of star formation in the case of $n_{\rm SF}$ = $0.1$~cm$^{-3}$. Since up to that moment, all models have experienced the same evolution, there is no difference between the red points in both panels. 
As can be seen in the left panel, the gas density in this N03 model reaches the star-formation threshold and star formation occurs. Moreover, the influence of supernova feedback can be seen in the green and blue points, where gas expands to larger radii and lower densities after having been heated. As is clear from the right panel, for $n_{\rm SF}$ = $6$~cm$^{-3}$ the gas simply keeps falling in. It will continue to do so during the first 4 Gyr until the built-up central pressure causes the gas to re-expand again. No stars are formed during the course of this simulation. 

\begin{figure}
 \centering \includegraphics[width=0.5\textwidth]{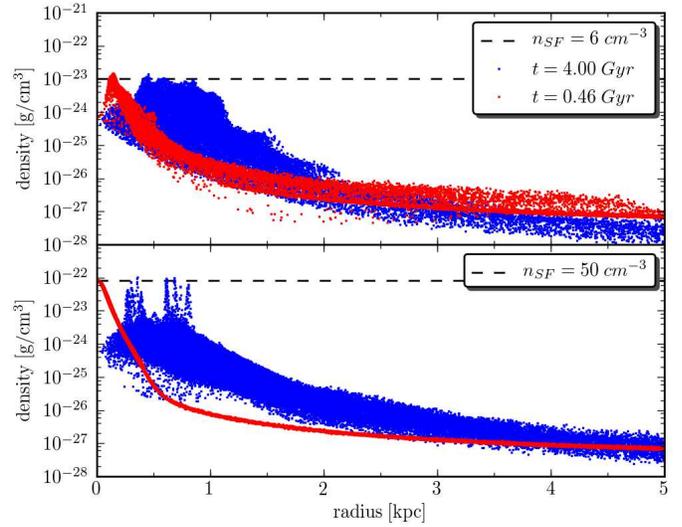}
 \caption{The density distribution of the ISM at different times for the N07 model, with different density thresholds and a fixed feedback efficiency of 0.7. \label{fig:N07}}
\end{figure}

As the density threshold is increased to higher values, star formation tends to occur more and more in small collapsed clumps. This becomes clear when comparing the panels from Figs. \ref{fig:N03} and \ref{fig:N07}. The latter shows the gas density distributions of two N07 models with $n_{\rm SF}$ = $6$~cm$^{-3}$ and $n_{\rm SF}$ = $50$~cm$^{-3}$. While the $n_{\rm SF}$ = $50$~cm$^{-3}$ model only exhibits star formation in a small number of discrete high-density clumps, the  $n_{\rm SF}$ = $6$~cm$^{-3}$ model lacks such well-defined clumps and star formation occurs more widespread.

\subsection{Scaling relations}
\label{section_scaling_relations}
In this section we discuss the properties of each of our models and draw some conclusions regarding the influence of the  $n_{\rm SF}$ and $\epsilon_{\rm FB}$ parameters on the models. An overview of some basic properties can be found in Table \ref{table_finalprop}.
\begin{table*}
 \begin{minipage}[ht]{\linewidth}
  \begin{center}
   \caption{Final properties of our large set of simulations. Columns: (1) model number (see Table \ref{Models_table}), (2) density treshold for star formation, (3) feedback efficiency, (4) final stellar mass, (5) starting time of star formation, (6) half-light radius, (7) mean surface \textbf{brightness} within the half-light radius, (8) central one dimensional velocity dispersion, (9) mass-weighted metallicity, (10) central surface brightness, (11) S\'ersic parameter, (12) circular velocity.} \label{table_finalprop} \begin{tabular}{lrrrrrrrrrrrrr}
\hline                                
$\mathrm{Model}$ & $\mathrm{n_{\rm SF}}$ & $\epsilon_{\rm FB}$ & $M_{\rm \star,f}$ & $\Delta T_{\rm SF}$ & $R_{\rm e}$ & $I_{\rm e}$ & $\sigma_{\rm 1D,c}$ & $V-I$ & $[Fe/H]$ & $\mu_{0}$ & $n$ & $V_{\rm c}$ \\ 
& $[cm^{-3}]$ & & $[10^{6}\ M_{\odot}]$ & $[Gyr]$ & $[kpc]$ & & $[km/s]$ & & & $[mag]$ & & $[km/s]$ \\ 
\hline
\hline
N03 & 0.1 & 0.1 & 0.285 & 0.342 & 0.100 & 3.176 & 6.806 & 0.835 & -1.183 & 23.603 & 1.496 & 16.176\\ 
\hline 
N05 & 0.1 & 0.1 & 5.667 & 0.168 & 0.230 & 10.779 & 12.190 & 0.860 & -1.236 & 23.170 & 0.959 & 20.396\\ 
\hdashline[1pt/5pt] 
N05 & 6 & 0.1 & 17.867 & 0.546 & 0.130 & 88.566 & 12.253 & 0.910 & -0.659 & 20.920 & 1.025 & 22.574\\ 
N05 & 6 & 0.3 & 4.049 & 0.546 & 0.142 & 18.222 & 9.131 & 0.870 & -1.130 & 23.005 & 0.877 & 20.483\\ 
N05 & 6 & 0.5 & 2.021 & 0.546 & 0.134 & 11.813 & 8.310 & 0.839 & -1.302 & 22.586 & 1.137 & 20.590\\ 
N05 & 6 & 0.7 & 1.174 & 0.546 & 0.118 & 9.252 & 8.107 & 0.817 & -1.542 & 22.694 & 1.231 & 20.555\\ 
N05 & 6 & 0.9 & 1.017 & 0.546 & 0.122 & 7.070 & 8.195 & 0.831 & -1.552 & 24.139 & 0.785 & 20.096\\ 
\hdashline[1pt/5pt] 
N05 & 50 & 0.3 & 6.116 & 0.688 & 0.244 & 10.945 & 8.346 & 0.852 & -1.016 & 23.154 & 1.104 & 21.158\\ 
N05 & 50 & 0.5 & 3.230 & 0.688 & 0.152 & 11.699 & 8.116 & 0.878 & -1.242 & 22.952 & 1.117 & 20.342\\ 
N05 & 50 & 0.7 & 2.128 & 0.688 & 0.141 & 10.182 & 8.362 & 0.829 & -1.426 & 23.141 & 1.112 & 20.038\\ 
N05 & 50 & 0.9 & 1.625 & 0.688 & 0.157 & 5.196 & 8.423 & 0.864 & -1.461 & 24.313 & 0.928 & 19.563\\ 
\hline 
N06 & 0.1 & 0.1 & 15.616 & 0.137 & 0.384 & 11.813 & 16.209 & 0.856 & -1.108 & 23.730 & 0.718 & 23.290\\ 
\hdashline[1pt/5pt] 
N06 & 6 & 0.1 & 42.542 & 0.460 & 0.150 & 198.791 & 16.779 & 0.870 & -0.540 & 20.383 & 0.892 & 27.277\\ 
N06 & 6 & 0.3 & 5.154 & 0.460 & 0.149 & 22.155 & 9.430 & 0.845 & -1.289 & 22.362 & 1.005 & 21.828\\ 
N06 & 6 & 0.5 & 3.425 & 0.460 & 0.156 & 16.053 & 8.875 & 0.832 & -1.329 & 22.981 & 0.872 & 21.335\\ 
N06 & 6 & 0.7 & 2.030 & 0.460 & 0.136 & 11.336 & 8.956 & 0.823 & -1.459 & 23.181 & 0.993 & 21.764\\ 
N06 & 6 & 0.9 & 2.255 & 0.460 & 0.161 & 9.701 & 8.640 & 0.830 & -1.437 & 23.493 & 0.935 & 21.668\\ 
\hdashline[1pt/5pt] 
N06 & 50 & 0.3 & 10.227 & 0.591 & 0.256 & 15.982 & 9.629 & 0.856 & -1.054 & 23.001 & 0.905 & 23.250\\ 
N06 & 50 & 0.5 & 5.780 & 0.591 & 0.173 & 16.660 & 9.040 & 0.866 & -1.228 & 22.497 & 1.142 & 21.697\\ 
N06 & 50 & 0.7 & 3.306 & 0.591 & 0.187 & 8.347 & 9.634 & 0.843 & -1.392 & 23.103 & 1.228 & 21.332\\ 
N06 & 50 & 0.9 & 2.718 & 0.591 & 0.180 & 7.722 & 9.041 & 0.838 & -1.408 & 23.093 & 1.228 & 21.089\\ 
\hline 
N07 & 0.1 & 0.1 & 67.575 & 0.135 & 0.693 & 14.994 & 23.992 & 0.887 & -0.808 & 23.281 & 0.889 & 30.289\\ 
\hdashline[1pt/5pt] 
N07 & 6 & 0.1 & 161.970 & 0.336 & 0.206 & 326.861 & 28.621 & 0.900 & -0.361 & 19.621 & 0.910 & 39.206\\ 
N07 & 6 & 0.3 & 14.933 & 0.336 & 0.220 & 31.447 & 10.274 & 0.843 & -1.133 & 21.839 & 1.076 & 23.908\\ 
N07 & 6 & 0.5 & 8.008 & 0.336 & 0.190 & 20.299 & 10.485 & 0.825 & -1.415 & 22.718 & 0.899 & 23.673\\ 
N07 & 6 & 0.7 & 5.046 & 0.336 & 0.192 & 13.261 & 9.642 & 0.816 & -1.480 & 22.874 & 1.054 & 23.759\\ 
N07 & 6 & 0.9 & 4.246 & 0.336 & 0.193 & 9.060 & 10.028 & 0.853 & -1.562 & 23.629 & 0.977 & 23.004\\ 
\hdashline[1pt/5pt] 
N07 & 50 & 0.3 & 21.037 & 0.460 & 0.322 & 18.593 & 9.452 & 0.870 & -1.056 & 22.234 & 1.146 & 24.773\\ 
N07 & 50 & 0.5 & 14.128 & 0.460 & 0.340 & 11.805 & 10.721 & 0.864 & -1.168 & 23.296 & 0.965 & 24.948\\ 
N07 & 50 & 0.7 & 9.027 & 0.460 & 0.477 & 3.644 & 10.586 & 0.862 & -1.294 & 24.391 & 1.066 & 24.018\\ 
N07 & 50 & 0.9 & 4.908 & 0.460 & 0.396 & 3.605 & 8.229 & 0.819 & -1.415 & 24.532 & 0.972 & 24.537\\ 
\hline 
N08 & 0.1 & 0.1 & 155.430 & 0.131 & 0.812 & 25.902 & 29.448 & 0.871 & -0.665 & 22.506 & 0.966 & 35.875\\ 
\hdashline[1pt/5pt] 
N08 & 6 & 0.1 & 271.070 & 0.278 & 0.163 & 839.934 & -99.000 & 0.893 & -0.261 & 17.300 & 1.467 & 43.269\\ 
N08 & 6 & 0.3 & 24.623 & 0.278 & 0.253 & 34.853 & 12.685 & 0.864 & -1.019 & 21.662 & 1.111 & 27.178\\ 
N08 & 6 & 0.5 & 12.423 & 0.278 & 0.248 & 17.704 & 12.089 & 0.838 & -1.404 & 22.986 & 0.846 & 25.561\\ 
N08 & 6 & 0.7 & 9.402 & 0.278 & 0.229 & 14.476 & 11.430 & 0.842 & -1.534 & 22.917 & 1.003 & 24.198\\ 
N08 & 6 & 0.9 & 0.610 & 0.278 & 0.086 & 128.570 & 8.947 & 0.454 & -4.277 & 20.621 & 1.077 & 27.395\\ 
\hdashline[1pt/5pt] 
N08 & 50 & 0.3 & 42.956 & 0.392 & 0.362 & 26.137 & 11.375 & 0.900 & -0.931 & 21.234 & 1.542 & 27.570\\ 
N08 & 50 & 0.5 & 22.743 & 0.393 & 0.481 & 9.039 & 11.535 & 0.860 & -1.147 & 23.855 & 0.768 & 27.479\\ 
N08 & 50 & 0.7 & 15.763 & 0.393 & 0.437 & 9.845 & 11.364 & 0.819 & -1.186 & 23.771 & 0.737 & 28.530\\ 
N08 & 50 & 0.9 & 9.019 & 0.393 & 0.400 & 5.897 & 12.584 & 0.842 & -1.331 & 24.193 & 0.836 & 27.050\\ 
\hline 
N09 & 0.1 & 0.1 & 394.500 & 0.104 & 0.616 & 109.262 & 38.883 & 0.841 & -0.382 & 20.759 & 0.943 & 48.060\\ 
\hdashline[1pt/5pt] 
N09 & 6 & 0.1 & 477.620 & 0.235 & 0.224 & 385.748 & -99.000 & 1.051 & -0.215 & 18.931 & 1.351 & 57.020\\ 
N09 & 6 & 0.3 & 86.095 & 0.235 & 0.278 & 82.718 & 18.482 & 0.898 & -0.911 & 20.249 & 1.388 & 31.210\\ 
N09 & 6 & 0.5 & 30.470 & 0.235 & 0.324 & 21.627 & 13.867 & 0.874 & -1.282 & 22.310 & 1.132 & 29.103\\ 
N09 & 6 & 0.7 & 19.274 & 0.235 & 0.500 & 6.815 & 11.576 & 0.843 & -1.356 & 23.629 & 0.969 & 29.240\\ 
N09 & 6 & 0.9 & 12.881 & 0.235 & 0.318 & 9.233 & 12.614 & 0.853 & -1.640 & 23.104 & 1.236 & 28.515\\ 
\hdashline[1pt/5pt] 
N09 & 50 & 0.3 & 94.102 & 0.324 & 0.382 & 45.638 & 14.780 & 0.901 & -0.917 & 20.402 & 1.541 & 31.697\\ 
N09 & 50 & 0.5 & 40.965 & 0.324 & 0.409 & 17.192 & 14.266 & 0.883 & -1.218 & 22.887 & 0.930 & 29.369\\ 
N09 & 50 & 0.7 & 23.972 & 0.324 & 0.559 & 5.967 & 13.400 & 0.867 & -1.330 & 23.974 & 1.037 & 29.311\\ 
N09 & 50 & 0.9 & 15.385 & 0.324 & 0.503 & 6.225 & 12.743 & 0.830 & -1.396 & 24.465 & 0.691 & 30.326\\ 
\hline 

\end{tabular}
 
  \end{center}
 \end{minipage}
\end{table*}
 
\subsubsection{Half-light radius $R_{e}$}
The half-light radius, or effective radius, denoted by $R_{e}$, encloses half of
a galaxy's luminosity. In panel a.) of Fig. \ref{fig:bigFig},
$R_{e}$ is plotted as a function of the $V$-band magnitude. 
The following trends can be observed in this figure:
\begin{itemize}
 \item For a fixed $n_{\rm SF}$, the effective radius varies only very slightly throughout the $\epsilon_{\rm FB}$-range and this without a clear trend between $R_{e}$ and $\epsilon_{\rm FB}$. However, for a fixed $n_{\rm SF}$ and dark-matter mass the stellar mass and consequently the luminosity decrease with increasing $\epsilon_{\rm FB}$. This is due to star formation being shut down more rapidly when feedback is more effective. As a result, galaxies tend to have higher stellar densities for smaller $\epsilon_{\rm FB}$.
 \item For a fixed $\epsilon_{\rm FB}$ of 0.1, an increase of $n_{\rm SF}$ from $0.1$ to $6$~cm$^{-3}$ results in a decrease of the effective radius. This is due to the size of the region where the SFC are fulfilled, which is much smaller for $n_{\rm SF}$ = $6$~cm$^{-3}$ than for $n_{\rm SF}$ = $0.1$~cm$^{-3}$, and the feedback is too weak to overcome this. In the case of an increase of $n_{\rm SF}$ from $6$ to $50$~cm$^{-3}$, the effective radius increases which is caused by the higher star formation peaks resulting in more supernovae explosions which redistribute the gas more efficiently.
 \item The simulations with high density threshold, $n_{\rm SF} > 0.1\ \rm cm^{-3}$, and high feedback efficiency, $\epsilon_{\rm FB} > 0.1$, have effective radii which are in agreement with the observations.
\end{itemize}
\textit{From this scaling relation we can constrain the $\epsilon_{\rm FB}$-parameter to be higher then 0.1 to produce galaxies with effective radii in agreement with observations of dwarf galaxies.}

\subsubsection{The fundamental plane}
\begin{figure*}
\begin{minipage}[ht]{\linewidth}
\begin{center}
 \centering 
 \includegraphics[width=0.95\textwidth,clip]{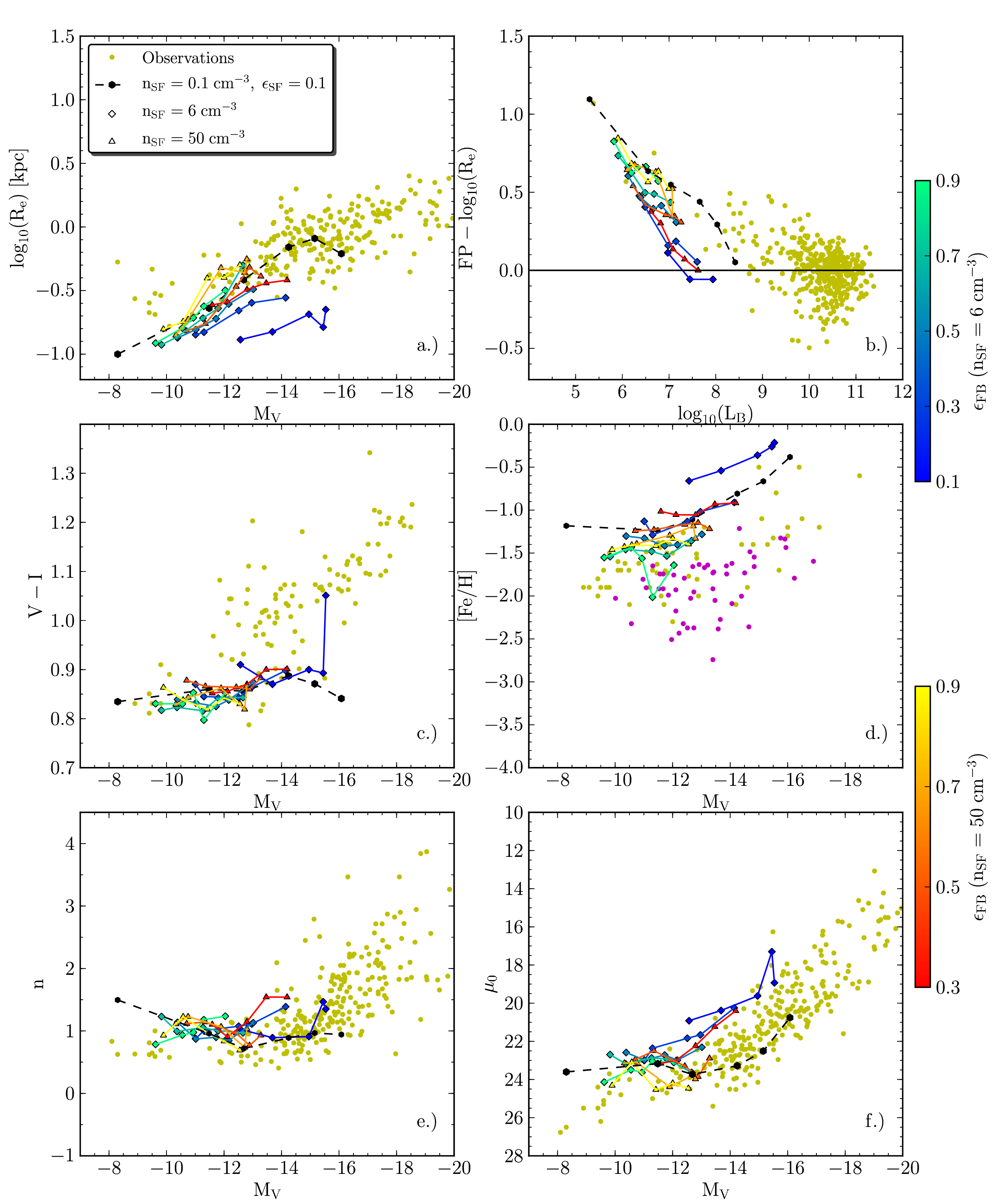}
 \caption{Some scaling relations and the surface brightness parameters as a function of the magnitude. In a.), the half-light radius $R_{\rm e}$ is plotted, b.) shows the vertical deviation of the simulated dwarf galaxies from the giant galaxies' FP, in c.) the $V-I$ color is plotted, d.) shows the iron content $[Fe/H]$. In panel e.) and f.), the S\'ersic index $n$ and central surface brightness $\mu_{0}$ are plotted. All these quantities are plotted against the $V$-band magnitude, except the FP which are plotted as a function of the $B$-band luminosity. The models with a density threshold of $6 \rm ~cm^{-3}$ and $50 \rm ~cm^{-3}$ are represented by blue-green diamonds and yellow-red triangles, respectively, where the colorscales represent a varying feedback efficiency. 
 For each color, the datapoints are connected by a line showing the mass evolution of the models. In the case of $n_{\rm SF}=0.1\ cm^{-3}$, represented by the black line, the models from N03 until N09 are plotted. In the cases of higher densities, represented by the colored lines, the datapoints are from models N05-N09.
 Our models are compared with observational data
   obtained from \citet{derijcke05}, \citet{graham03}
   , LG data come from \citet{peletier93},
   \citet{peletier93}, \citet{irwin95}, \citet{saviane96},
   \citet{grebel03}, \citet{mcconnachie06},
   \citet{mcconnachie07}, \citet{zucker07}, Perseus data from
   \citet{derijcke09}, Antlia data from
   \citet{castelli08}. For the $[Fe/H]-M_{V}$ plot, data from \citet{grebel03}, \citet{sharina08} and \citet{lianou10} was used, the yellow and magenta dots represent data from dSph and dIrr galaxies, respectively. \label{fig:bigFig}}
 \end{center}
 \end{minipage}
\end{figure*}

The fundamental plane (FP) is an observed relation between the effective radius, $R_{\rm e}$, the mean surface brightness within the effective radius, $I_{\rm e}$, and the central velocity dispersion, $\sigma_{\rm c}$ of giant elliptical galaxies. It is a linear relation, given by
\begin{equation}
 \log( R_{\rm e} ) = -0.629 - 0.845 \log( I_{\rm e} ) + 1.38 \log( \sigma_{\rm c} ),
\end{equation}
between the logarithms of these quantities \citep{burstein97}. In panel b.) of Fig. \ref{fig:bigFig}, we plot the ''vertical'' deviation  of the simulated galaxies from the giant galaxies' FP.

Dwarf galaxies generally lie above the FP in this projection. This is thought to be a consequence of their having shallower gravitational potential wells than giant galaxies. This, together with
the feedback, results in more diffuse systems. Models with a high star-formation threshold in combination with a low supernova feedback turn out to be very compact. They actually populate
the FP at low luminosities. However, this region of the three-dimensional space spanned by $\log(R_{\rm e})$, $\log( I_{\rm e} )$, and $\log( \sigma_{\rm c} )$ is observed to be devoid of galaxies.
Hence, models with low stellar feedback, $\epsilon_{\rm FB}$ up to $0.3 $, and high density thresholds, $n_{\rm SF} > 0.1$~cm$^{-3}$, can be rejected.

\subsubsection{Color $V-I$}
Fig. \ref{fig:bigFig}, panel c.) shows the $V-I$ color in function of the $V$-band magnitude. 
The color scatter between the different models is rather small.  The observed galaxies follow a mass-metallicity relation so the metallicity generally increases with the galaxy (stellar) mass, resulting in increasing $V-I$ values for increased galaxy mass. Within the relatively small mass range covered by the models, color
is only a very weak function of stellar mass. For a fixed feedback efficiency, when increasing the density threshold the $V-I$ also increases slightly resulting in bluer galaxies for the models with low density threshold. This is due to the effect that stars are formed in more metal enriched regions in the models with high density threshold. 
When the density threshold is kept constant and only the feedback efficiency is increased the $V-I$ slightly decreases, so the models get slightly bluer due to a dilution of the gas when it is more spread out by supernovae explosions.

\subsubsection{Metallicity}
In  panel d.) of Fig. \ref{fig:bigFig} a plot of iron content [Fe/H] as a function of the $V$-band magnitude is shown. The mass-weighted value of [Fe/H] is a measure of the metallicity of a galaxy. The yellow and magenta dots represent observational data from  dwarf spheroidal and dwarf elliptical galaxies and irregular dwarf galaxies, respectively. 
Some general conclusions we can take away from this figure are:
\begin{itemize}
 \item Low-mass models with low density threshold, $n_{\rm SF} \approx 0.1$~cm$^{-3}$, and low feedback, $\epsilon_{\rm FB} \approx 0.1$, keep forming stars throughout cosmic history 
and do not expel enriched gas. As a consequence, they turn out to be too metal rich, compared with observed dwarf galaxies. Models with higher $n_{\rm SF}$ compare much 
more favorably with the data in this respect.
  \item For a fixed $n_{\rm SF}$, increasing $\epsilon_{\rm FB}$, produces more metal poor galaxies. This is likely due to the fact that the increased feedback 
  extinguishes star formation more rapidly and disperses the metal enriched gas more widely.
 \item Increasing $n_{\rm SF}$ at fixed $\epsilon_{\rm FB}$ and fixed mass, results in an increase of the metallicity and of the stellar mass when going from $n_{\rm SF} = 0.1$~cm$^{-3}$ to $n_{\rm SF} = 6$~cm$^{-3}$. A further increase of $n_{\rm SF}$ at fixed $\epsilon_{\rm FB}$, up to $n_{\rm SF} = 50$~cm$^{-3}$, has a much smaller impact on metallicity and stellar mass. The former is likely due to more vigorous star formation in less easily dispersable high density regions. 
 \end{itemize}

\subsubsection{Surface brightness profiles}
We fitted a S\'ersic profile, of the form
\begin{equation}
 I(R)=I_{0}e^{-\left(\frac{R}{R_{0}}\right)^{1/n}},
\end{equation}
to the surface brightness profiles of the simulated galaxies.
The S\'ersic parameter $n$ and the central surface brightness $\mu_{0}$ are plotted respectively in the panels e.) and f.) of Fig. \ref{fig:bigFig} as a function of the $V$-band magnitude.
\begin{itemize}
 \item For a fixed $n_{\rm SF}$, when increasing the $\epsilon_{\rm FB}$, there is a weak trend for the S\'ersic parameter $n$ and the central surface brightness to decrease. More vigorous
feedback appears to result in more diffuse dwarf galaxies, as one would expect.
 \item As an echo of the $R_{\rm e}-M_{\rm V}$ relation, simulations with high density threshold, $n_{\rm SF} > 0.1\ \rm cm^{-3}$, and low feedback efficiency, $\epsilon_{\rm FB} = 0.1-0.3$, are systematically too compact, with $\mu_0 \sim 20$~mag~arcsec$^{-2}$, compared with the observations.
 \item The models with high density thresholds and strong feedback are in general agreement with the observations.
\end{itemize}

\subsubsection{The Tully-Fisher relation}
\begin{figure}
 \centering \includegraphics[width=\columnwidth]{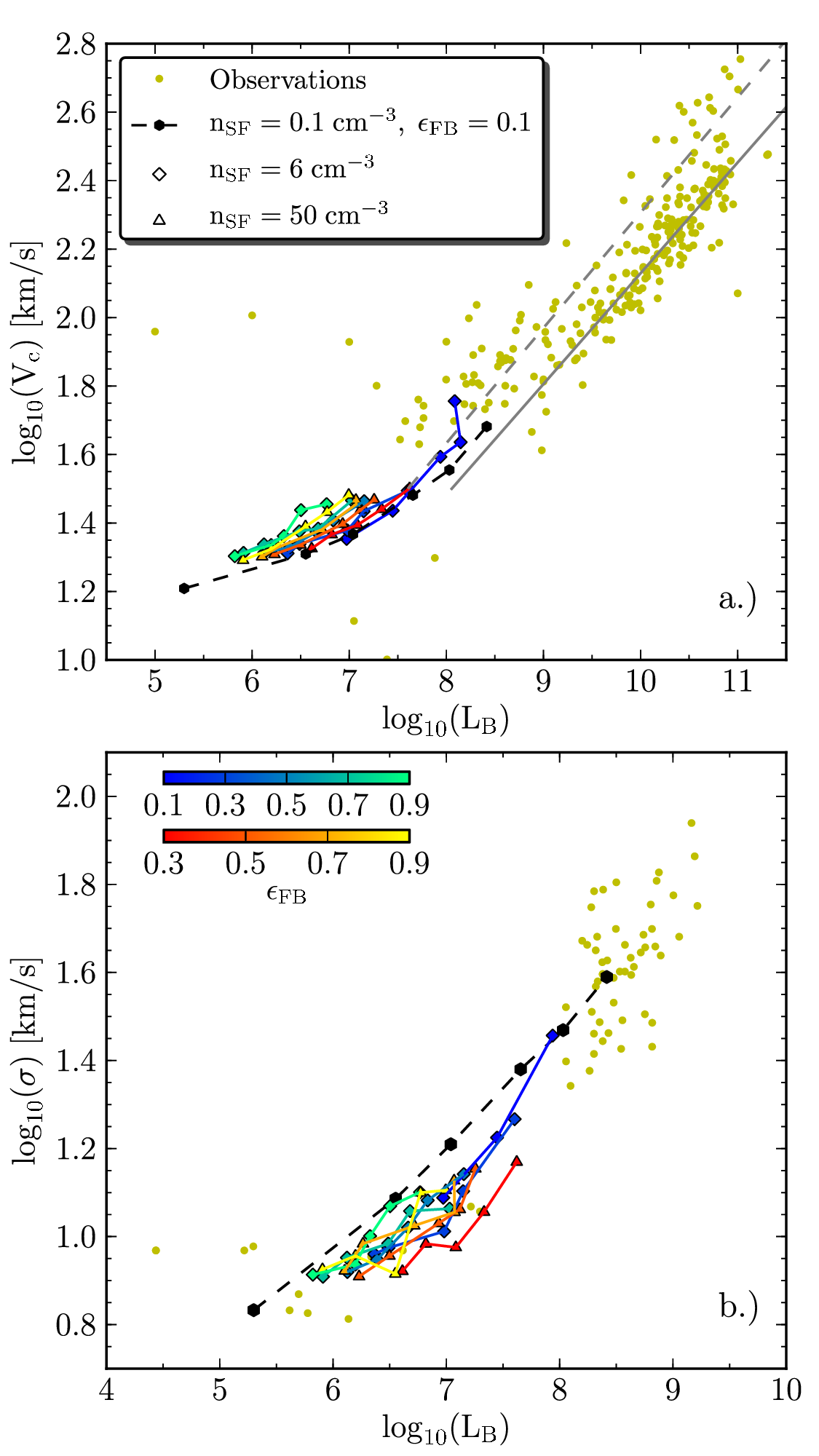}
 \caption{The top panel shows the Tully-Fisher relation between the circular
   velocity and the luminosity in the B-band. The full gray line shows the TF relation for early type galaxies, the dashed gray line is the TF relation of spiral galaxies as determined by \citet{derijcke07}. The lower panel shows the Faber-Jackson relation between the velocity
   dispersion and the luminosity in the B-band.}  
 \label{fig:TF_FJ}
\end{figure}

Panel a.) of Fig. \ref{fig:TF_FJ} shows the B-band Tully-Fisher
relation (TFR) between the circular velocity, denoted by $V_{c}$, and the luminosity in the B-band, $L_{\rm B}$. The simulations are compared with observational data and with the Tully-Fisher relation for early-type (full gray line) and for spiral galaxies (dotted gray line) that was determined by \cite{derijcke07}. 
All simulations predict that the TFR becomes substantially shallower in the dwarf regime, below luminosities of the order of $L_{\rm B} \sim 10^7\,L_{\odot, \rm B}$.  This can be seen as a consequence of the
very steep $M_{\rm star}-M_{\rm halo}$ relation in the dwarf galaxy regime (see paragraph \ref{subsubMstMh}). For a fixed $n_{\rm SF}$, an increase in feedback efficiency does not influence $V_{c}$ very much since there are so few stars that $V_c$ is set by the dark-matter halo. The effect on the stellar mass, and consequently on $L_{\rm B}$, is, however,
quite large. Therefore, increasing $\epsilon_{\rm FB}$ at fixed $n_{\rm SF}$ and dark-matter mass causes galaxies to shift leftwards in panel a.) of Fig. \ref{fig:TF_FJ}. Except for this effect, once $n_{\rm SF}$ and $\epsilon_{\rm FB}$ are raised above their minimum values of $0.1$~cm$^{-3}$ and $0.1$, respectively, there is no significant differences between the TFRs traced by the different series of models.

\subsubsection{The Faber-Jackson relation}
The Faber-Jackson (FJR) relation, plotted in panel b.) of Fig. \ref{fig:TF_FJ} is the relation between the stellar central velocity dispersion and
the luminosity in the $B$-band. The stellar central velocity dispersion is a projection of the velocity dispersion along the line of
sight. This is measured by fitting an exponential function to the dispersion profile and retaining the maximum of the function as the
central value. 

From this figure we see:
\begin{itemize}
 \item For a fixed $n_{\rm SF}$, when increasing the $\epsilon_{\rm FB}$, the velocity dispersion decreases first after which it settles around a value which depends on the dark-matter mass of the model.
 \item For a fixed $\epsilon_{\rm FB}$, when increasing $n_{\rm SF}$, only a minor influence on the velocity dispersion is observed.
\end{itemize}

\subsubsection{The $M_{\rm star}$-$M_{\rm halo}$ relation} \label{subsubMstMh}
\begin{figure}
 \centering
 \includegraphics[width=\columnwidth]{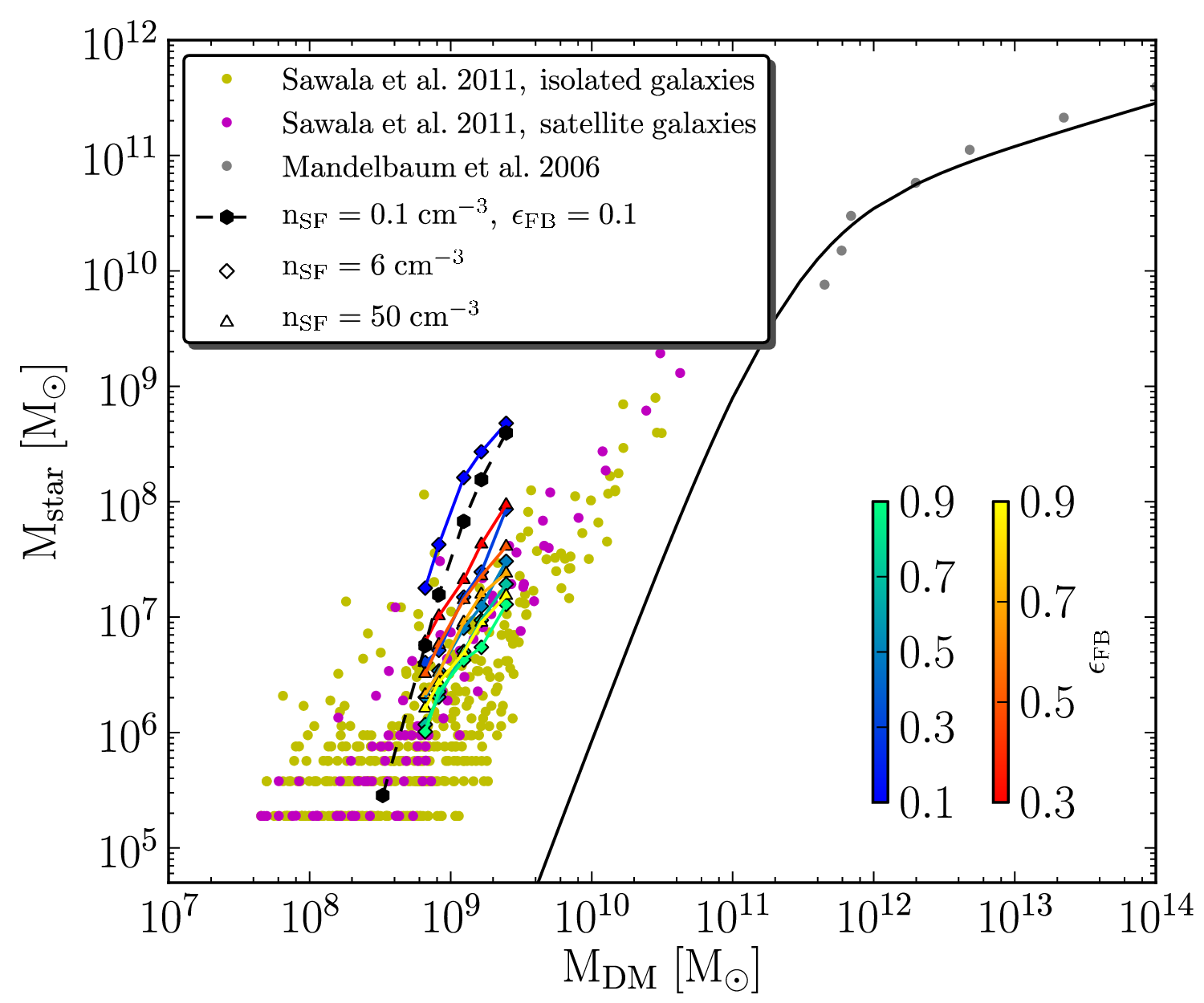}
 \caption{ The stellar mass versus the DM halo mass, plotted in
   comparison with the models by \citet{sawala11b}. The gray dots show data from gravitational lensing from \citet{mandelbaum06}. The black
   line is the trend for this relation that was determined by
   \citet{guo10}. \label{fig:MstarMhalo}}
\end{figure}

In Fig. \ref{fig:MstarMhalo}, the $M_{\rm star}$-$M_{\rm halo}$ relation of the simulations at $z=0$ is plotted. 
We can make similar conclusions here as were made in the SFH section:
\begin{itemize}
 \item If $n_{\rm SF}$ is fixed, the stellar mass will decrease if the $\epsilon_{\rm FB}$ is increased. This is what was expected because with more
feedback the gas is distributed over a larger area and
the infall of the gas to the appropriate density threshold will take
longer.
 \item If $\epsilon_{\rm FB}$ is fixed, for increasing $n_{\rm SF}$, the stellar mass increases too. When feedback is very small, the gas density will stay high and the star formation will not be interrupted, resulting in a high stellar mass. The effect is smaller for higher feedback.
\end{itemize}

In Fig. \ref{fig:MstarMhalo}, our different sets of
models are found to be in agreement with the results from
the Aquila simulation where a density threshold of 10 $\rm cm^{-3}$ and a feedback efficiency of 0.7 was used. 
While the initial conditions of our dwarf galaxy simulations are admittedly quite simplified, they do have high spatial resolution
and realistic implemented physics. It is therefore encouraging that they compare favorably with cosmological simulations 
like the Aquila simulation, which have cosmologically well motivated initial conditions but in which dwarf galaxies are very close to the resolution limit \citep{sawala11b}.
However it is impossible by further tuning of the feedback efficiency and/or the density threshold to reproduce the trend that was derived by \cite{guo10}. 

By increasing the density threshold and feedback efficiency, the stellar mass is reduced by almost two orders of magnitude, but there still remains a difference of many orders of magnitude between our simulations and the $M_{\star}$-$M_{halo}$ relation from \cite{guo10}. It is also interesting to notice that although our models do not reproduce the relation, they do have a very similar slope.

\section{Discussion and conclusions}
\label{section:results}

\subsection{Cusp to core}
Whether the halo density profile is cusped or cored has been a point of discussion for quite some time. Observationally, evidence for cored DM profiles is found \citep{gentile04}, but from cosmological DM simulations a cusped density profile is deduced \citep{NFW, moore96}. The inherent limitation due to the angular resolution of the observations is ruled as a cause of the observed flat density profiles by \cite{deblok02}. \cite{gentile05} also excluded the possibility of non-circular gas motions which might result in a rotation curve that is best fitted by a cored halo, while the dark matter halo actually has a cuspy profile. However, from the simulation point of view, \cite{mashchenko06} mentioned a natural transition of a cusp to a flattened core when the dark matter halo is gravitationally heated by bulk gas motions. 

Our simulations are set up with a cusped NFW halo in agreement with cosmological simulations. The infall of gas causes an adiabatic compression of the dark halo. When gas is evacuated from the central 
regions, be it by a fast re-expansion as the gas pressure builds up or by supernova feedback, the dark-matter halo reacts non-adiabatically and kinetic energy of the gas is transferred to the dark matter. This results in a flattening of the central density and so the cusp is converted into a core. We can conclude that the conversion of the cusped halo density profile to a cored profile is realized by the removal of baryons
from the galaxy center \citep{read05}, whether this is due to a re-expansion of the gas or by feedback effects or by another process.

\subsection{Degeneracy}
By increasing both the  density threshold and the feedback efficiency, the simulated galaxies  move along the observed kinematic and photometric scaling relations. 
These two parameters, the feedback efficiency $\epsilon_{\rm FB}$ and the density threshold $n_{\rm SF}$, correlate with each other and an increase of the one can be counteracted by an increase of the other, resulting in galaxies with similar properties. To be more specific:~the individual galaxies are drastically different for different parameter values but they all line up along the same scaling relations and 
can therefore be seen as good analogs of observed dwarf galaxies. 

The feedback efficiency quantifies the fraction of the $10^{51}$~ergs of energy that are released during a supernova explosion and thermally injected into the ISM. For each value of the density threshold we can determine the feedback efficiency range for which the models are in agreement with the observations, although we are not able to deduce a unique $n_{\rm SF}$/$\epsilon_{\rm FB}$-combination which would be the ``correct`` representation of the physical processes that happen in galaxies. 

For a certain density threshold, a lower limit of the corresponding $\epsilon_{\rm FB}$-parameter can be determined from the effective radius: the galaxies become too centrally concentrated when the feedback is too low. 
From the scaling relations we cannot deduce an upper limit for the $\epsilon_{\rm FB}$-parameter, but one could argue that the ISM cannot receive more energy than there is released by the supernova explosion, resulting in a maximal value for the feedback efficiency of 1.

In the case of a density threshold of $n_{\rm SF}=0.1$~cm$^{-3}$, the models are generally in good agreement with the observations besides the somewhat high metallicities. This is also the reason why the feedback efficiency was not varied in this case. If we compare the high density threshold models, $n_{\rm SF}>0.1$~cm$^{-3}$, with the observations we can conclude that the feedback efficiency should be larger then $\sim 0.3$. For a density threshold of $n_{\rm SF}=6$~cm$^{-3}$, we prefer a value of 0.7 for the feedback. Similarly we prefer a feedback efficiency of 0.9 in the case of a density threshold of $n_{\rm SF}=50$~cm$^{-3}$

The fact that different $n_{\rm SF}$/$\epsilon_{\rm FB}$-combinations result in simulated galaxies with properties that are in agreement with the observations invokes a warning for future simulations and indicates that there is still some work left to determine the density of the star forming regions and the fraction of supernova energy that is absorbed by the ISM, quantities which are hard to determine observationally. 

There are however other parameters that might influence the starformation rate and our degeneracy, which are not investigated here:
\itemize{ 
  \item{Given the fact that the star-formation efficiency $c_\star$ was found by other authors not to have a significant impact on stellar mass, we did not investigate it in detail in this paper.}
  \item{The choice of the IMF, for which in our simulations a Salpeter IMF is used, determines the mass distribution of stars. The fraction of high-mass stars  influences the number of SNIa and SNII explosions and as a consequence it will influence the amount of feedback and the chemical evolution. However, given 
the large number of IMF parameterizations available in the literature, testing them is a very daunting task which falls outside the scope of this paper. Moreover,
part of the IMF-variation is quantified approximately by the variation in $\epsilon_{\rm FB}$ which we do investigate. }
  \item{There are other possible feedback implementations, next to the release of feedback energy as thermal energy to the gas. It also could be released as kinetic energy by kicking the gas particles or by blast-wave feedback \citep{mayer08}.}
\item Other implementations of star formation, e.g. based on a subgrid model of H$_2$-formation \citep{pelupessy04}, are possible. 
}


\subsection{The dwarf galaxy dark-matter halo occupancy}
\begin{figure}
 \centering \includegraphics[width=\columnwidth]{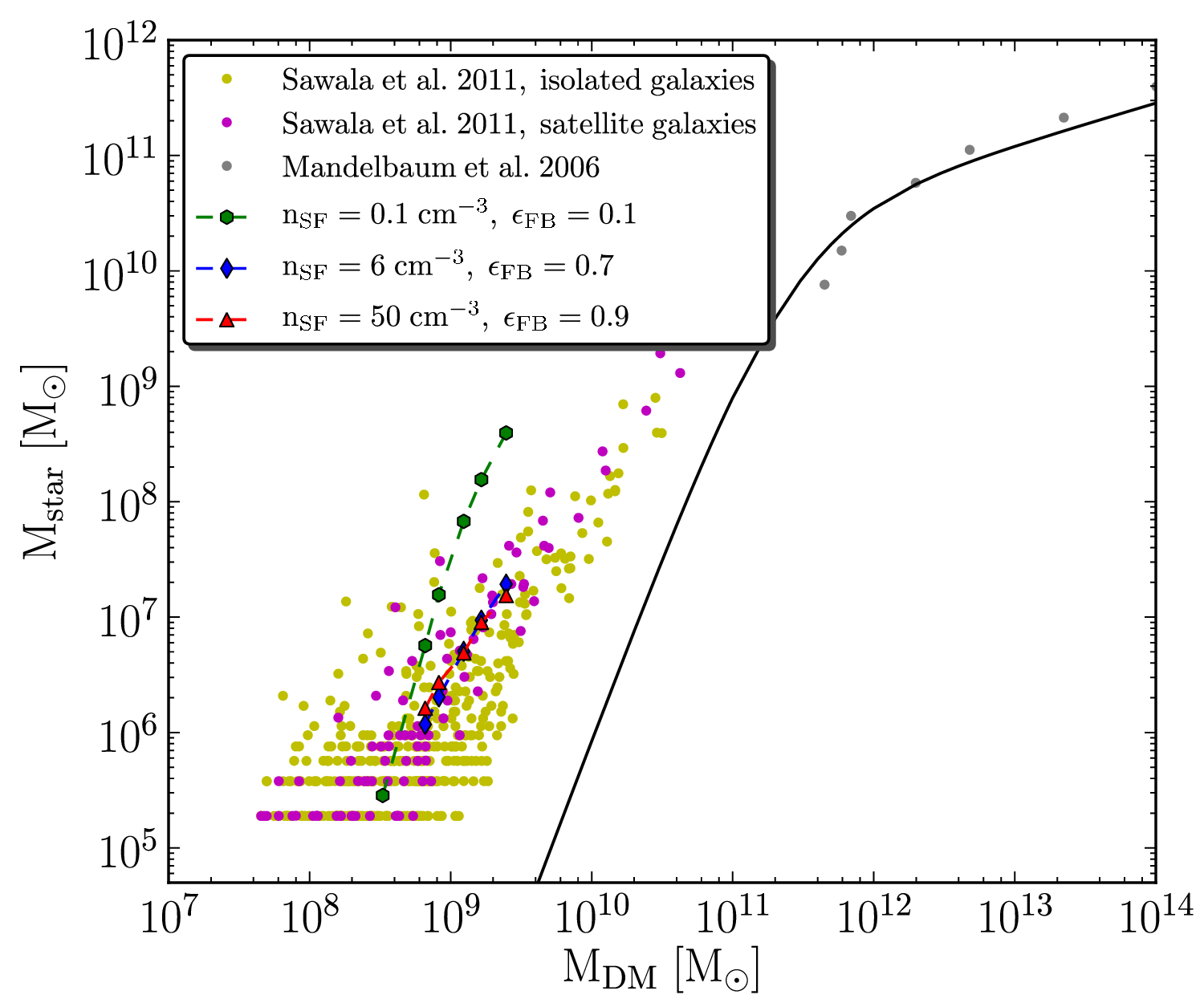}
 \caption{The $M_{\rm star}-M_{\rm halo}$ of our best models for different density threshold compared to the relation of \citealt{guo10}, other simulations from \citealt{sawala11b} and observations from \citealt{mandelbaum06}.
 \label{fig:Mst_Mh_f}}
\end{figure}

To conclude, Fig. \ref{fig:Mst_Mh_f} shows the models which best agree with the observations for each density threshold that was used in our analysis.
Increasing $n_{\rm SF}$ together with $\epsilon_{\rm FB}$ leads to a strong reduction, of almost two orders of magnitude, of the stellar mass, especially in the most massive models. 
However, with the physics included in our simulations, we are unable to reproduce the $M_{\rm star}-M_{\rm halo}$ relation of \cite{guo10}. Surprisingly, the best models trace a $M_{\rm star}-M_{\rm halo}$ relation
with a slope that is similar to that of the relation of \cite{guo10}. Our simulations are in agreement with results from  cosmological simulations, which have, however, much lower spatial resolution in the dwarf regime \cite{sawala11b}. We did not explore yet higher values for $n_{\rm SF}$ and $\epsilon_{\rm FB}$ because it is clear from Fig. \ref{fig:Mst_Mh_f} that the reduction of $M_{\rm star}$ stagnates for
high $n_{\rm SF}$-values. Moreover, to compensate for the high density threshold, an unphysical large value for $\epsilon_{\rm FB}$, higher than 1, would be required. Thus, we arrive at $(n_{\rm SF}=6\,{\rm cm}^{-3},\, \epsilon_{\rm FB}\sim 0.7)$ and $(n_{\rm SF}=50\,{\rm cm}^{-3},\, \epsilon_{\rm FB}\sim 0.9)$
as the models which are in best agreement with the observed photometric and kinematical scaling relations and with the $M_{\rm star}-M_{\rm halo}$ relation derived directly from cosmological simulations.

While it appears impossible to place isolated dwarf galaxies on the $M_{\rm star}-M_{\rm halo}$ relation of \citet{guo10}, it is possible to envisage external influences that may further reduce
$M_{\rm star}$, as already mentioned in the Introduction:
\begin{itemize}
 \item Not properly taking into
  account the effects of reionisation may lead to an overestimation of
  the gas content of dwarfs and an underestimation of the gas cooling
  time. However, even taking into account reionisation, the dwarf galaxies simulated by \citet{sawala11a} 
  had much too high stellar masses.
\item At a given gas density, the star-formation efficiency of dwarf galaxies could be lower than that of more massive stellar systems because of their lower metallicity and hence lower dust content.
  This could be mimicked by reducing the star-formation efficiency parameter $c_\star$ (see eq. (\ref{cstar})) in the dwarf regime. However, \citet{stinson06} have shown that, because of self-regulation, the star-formation rate is very insensitive to this parameter:~varying $c_\star$ between 0.05 and 1 left the mean star-formation rate virtually unchanged.
\item External processes such as ram-pressure stripping and tidal stirring may lead to a premature cessation of star formation and hence lower $M_{\rm star}$ \citep{mayer06}. However, these processes are
only effective if the gravitational potential wells of dwarf galaxies are sufficiently shallow and if they are stripped early enough in cosmic history, before they converted their gas into stars. It is
unclear whether these constraints are met. In \citet{derijcke10}, and references therein, it was argued that the number of red-sequence, quenched dwarf galaxies increased significantly over the last half of the Hubble time and that the dwarf galaxies now residing in the Fornax cluster were accreted less than a few crossing times age (i.e. less than a few Gyr). This timescale would have left dwarf galaxies ample time to form stars before entering the cluster.
\end{itemize}

\section*{Acknowledgements}
We thank Volker Springel for
making publicly available the {\sc Gadget-2} simulation code and Till Sawala for making available to us his
data from the Aquila simulation. We also would like to thank the anonymous referee for his/her stimulating remarks
that have greatly improved the manuscript.





\label{lastpage}

\end{document}